\documentclass[aps,twocolumn,nofootinbib,groupedaddress,amsfonts,floatfix]{revtex4-1} 
\usepackage{graphicx,amsmath,amssymb,amstext}
\usepackage{amssymb,amsbsy,amsfonts,amsthm,color}

\usepackage{epsfig}
\usepackage{graphicx}
\usepackage{subfigure}
\usepackage{hyperref}
\usepackage{cancel}

\graphicspath{{Figures/}}

\usepackage{color}
\usepackage[dvipsnames]{xcolor}

\newcommand{\Beq}{\begin{eqnarray}}
\newcommand{\Eeq}{\end{eqnarray}}

\newcommand{\nn}{\nonumber \\}

\def\lsim{\mathrel {\vcenter {\baselineskip 0pt \kern 0pt \hbox{$<$} \kern 0pt \hbox{$\sim$} }}}

\newcommand{\mpl}{M_{\mbox{\tiny Pl}}}

\def\gsim{\mathrel {\vcenter {\baselineskip 0pt \kern 0pt \hbox{$>$} \kern 0pt \hbox{$\sim$} }}}
\def\p{\partial}

\newcommand{\eqn}[1]{eq. \ref{#1}}

\begin{document}
{\hfill KCL-PH-TH/2018-44}
\title{Cosmic String Loop Collapse in Full General Relativity}

\author{Thomas Helfer${^a}$}
\email{thomashelfer@live.de}
\author{Josu C. Aurrekoetxea ${^a}$}
\email{j.c.aurrekoetxea@gmail.com}
\author{Eugene A. Lim${^a}$}
\email{eugene.a.lim@gmail.com}

\affiliation{${^a}$Theoretical Particle Physics and Cosmology Group, Physics Department,
Kings College London, Strand, London WC2R 2LS, United Kingdom}
\begin{abstract}
We present the  first fully general relativistic dynamical simulations of Abelian Higgs cosmic strings using 3+1D numerical relativity. Focusing on cosmic string loops, we show that they collapse due to their tension and can either (i) unwind and disperse or (ii) form a black hole, depending on their tension $G\mu$ and initial radius. We show that these results can be predicted using an approximate formula derived using the hoop conjecture, and argue that it is independent of field interactions. We extract the gravitational waveform produced in the black hole formation case and show that it is dominated by the $l=2$ and $m=0$ mode.  We also compute the total gravitational wave energy emitted during such a collapse, being $0.5\pm 0.2~ \%$ of the initial total cosmic string loop mass, for a string tension of $G\mu=1.6\times 10^{-2}$ and radius $R=100~\mpl^{-1}$. We use our results to put a bound on the production rate of planar cosmic strings loops as $N\lsim 10^{-2}~\mathrm{Gpc}^{-3}~\mathrm{yr}^{-1}$.
\end{abstract}
\maketitle

\section{Introduction} \label{sect:intro}
The recent detection of Gravitational Waves (GW) from black hole (BH) \cite{Abbott:2016blz} binaries by the LIGO/VIRGO collaboration marked the start of a new era of observations. Beyond astrophysical objects such as BH and neutron stars, this paved the way for the use of GW to search directly for signatures of new physics. One of the key targets for this search are cosmic strings \cite{Kibble:1976sj,Abbott:2017mem,Copeland:2009ga}.

Cosmologically, cosmic strings networks naturally arise after a phase transition in the early universe, possibly during GUT symmetry breaking.  More speculatively, string theory also suggests the presence of cosmological fundamental superstrings, especially through the mechanism of brane inflation \cite{Pogosian:2003mz,Jones:2003da}.  These networks may  manifest themselves through several channels, such as imprints via lensing on the Cosmic Microwave Background \cite{Ade:2013xla} and possibly through the presence of a stochastic gravitational wave background. The latter in particular is recently searched for by the LIGO/VIRGO collaboration \cite{Abbott:2017mem}. More intriguingly, one can search for localized coherent events of these strings, such as when the strings self-interact through the formation of sharp cusps or through the collisions of traveling kinks that are formed during the intercommutation (i.e. collisions) of cosmic strings. 

Before this work, the two primary methods of modeling cosmic strings has been through solving the field theory equations in flat or expanding spacetime, or through an effective Nambu-Goto prescription with weak coupling to gravity  (see e.g. \cite{Vilenkin:2000jqa}). In either case, by considering the stress-energy of a network of strings, one can then compute in the weak gravity limit a stochastic GW background \cite{Vilenkin:1981bx,Damour:2004kw}.   Local events such as the collisions of traveling kinks and cusps along the strings are expected to produce bursts of GW -- these bursts events have been computed using the Nambu-Goto approximation, again in the weak field limit \cite{Damour:2004kw}. These two methods do not coincide in general, mainly due to their disagreement on the primary energy loss mechanism of the cosmic strings (see \cite{Hindmarsh:2017qff,Vincent:1997cx,Moore:1998gp,Olum:1998ag, Moore:2001px,BlancoPillado:2011dq}). 
\begin{figure}[tb]
\begin{center}
{\includegraphics[width=0.9\columnwidth]{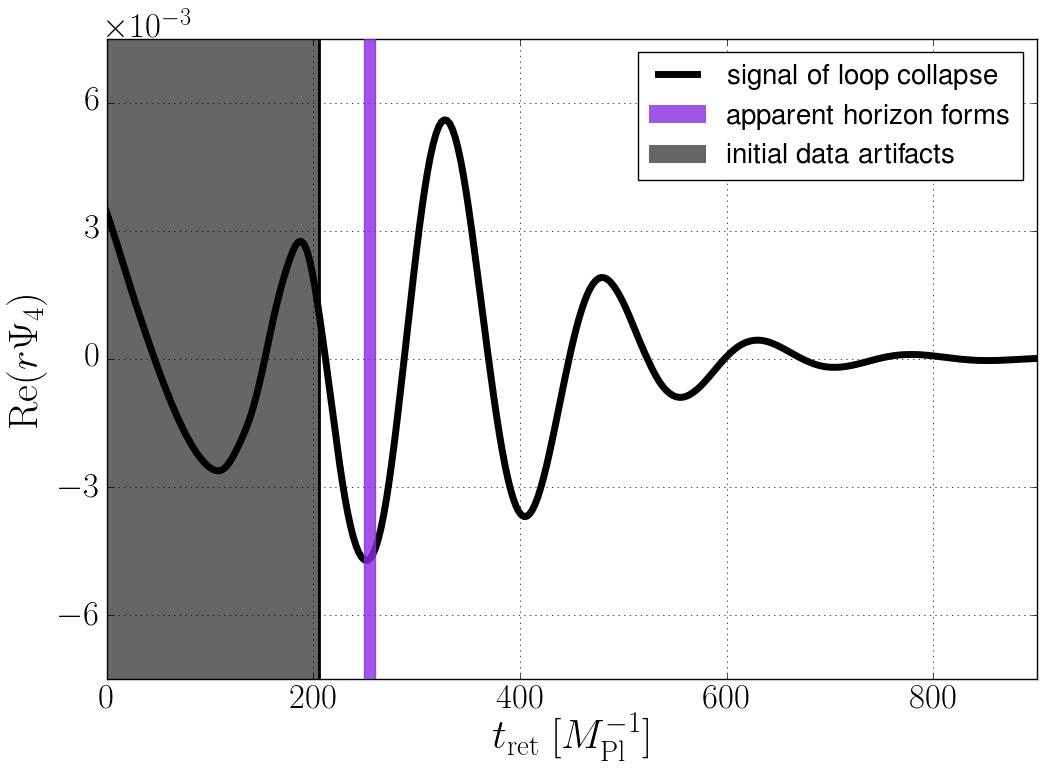}}
\caption{\textbf{GW for a BH formed from circular cosmic string loop collapse:} We plot the real part of the dominant $l = 2$ $m = 0$ mode of  $r\Psi_4$ over time. The loop has tension $G\mu = 1.6\times 10^{-2}$ and an initial radius $R = 100~\mpl^{-1}$. The grey shaded area of the plot are mixed with stray GWs that arise as artifacts of the initial data. The x-axis $t_{\mathrm{ret}}=t-r_\mathrm{ext}$ is the retarded time where $r_\mathrm{ext}$ is the extraction radius.}  
 \label{fig:WaveForm_sub}
\end{center}
\end{figure}

Going beyond the weak field limit requires the finding of the solutions to the full field theory coupled to general relativity -- and in this work we present the \emph{first numerical relativity simulation of Abelian Higgs cosmic strings in full general relativity}. In this first paper of a series, we numerically explore the collapse of a circular cosmic string loop in extreme regimes ($4\times 10^{-3}<G\mu<4\times 10^{-2}$). We show that whether the loop collapses into a BH or unwinds itself depends on a simple analytic relation derived using the hoop conjecture. In the former case, we computed both the gravitational waveform (fig. \ref{fig:WaveForm_sub}) and its integrated GW energy emitted from such a collapse. For the latter, we found that the total energy emitted in gravitational waves is $0.5\pm 0.2~\%$ of the initial mass, which is in agreement with the bound of $<29\%$ \cite{HAWKING199036}.We will discuss direct detection prospects of such individual collapse events with GW detectors in section \ref{sect:discussion}.


\section{Abelian Higgs with Gravity} \label{sect:numset}
The action of the Abelian Higgs model minimally coupled to gravity \footnote{We use the $-+++$ convention for the metric, and set $\hbar=c=1$ and $\mpl=1/\sqrt{G}$.}
\begin{equation}
S=S_{EH}-\int d^4x\sqrt{-g}\left[(D_{\mu}{\phi})^{*}(D^{\mu}{\phi})+\frac{1}{4}F_{\mu\nu}F^{\mu\nu}+V(\phi) \right]~,
\end{equation}
where $S_{EH} = \int dx^4 \sqrt{-g}(R/16\pi G)$, $D_{\mu}$ is the covariant derivative $(\partial_{\mu}-ieA_{\mu})$ with its $U(1)$ gauge field $A^{\mu}$, and $V(\phi)$ is the potential of the complex scalar field $\phi$  given by
\begin{equation}
V(\phi) = \frac{1}{4}\lambda \left(\left|\phi\right|^2-\eta ^2\right)^2~,~ F_{\mu\nu} = \partial_{\mu}A_{\nu}-\partial_{\nu}A_{\mu}~.
\end{equation}
For simplicity, we set the charge $e$ and the dimensionless coupling constant $\lambda$ to obey the critical coupling limit
\begin{equation}\label{eqn:critcoup}
\beta = \frac{\lambda}{2e^2} = 1~,
\end{equation}
in which the Higgs and vector masses are identical and $\mu$ simplifies to 
\begin{equation}\label{eqn:Gmu}
\mu = 2\pi\eta^2~.
\end{equation}

As a check of our code,  we numerically construct a fully relativistic infinite static string coupled to gravity and demonstrate that its evolution is indeed static and stable. The details of this construction can be found in Appendix \ref{sect:Static String}. 

In this paper, we consider circular string loops. To construct the initial conditions, we define toroidal coordinates 
\begin{equation}
\begin{split}
x&=\cos\varphi (R+r \cos\theta )~, \\
y&=\sin\varphi (R+r \cos\theta )~, \\
z&=r \sin\theta ~, \\
\end{split}
\end{equation}
where $R$ is the radius of the loop and choose the following ansatz for the field variables 
\begin{equation}
\phi = f(r)e^{in\theta}~,~A_{\theta} = \frac{n\alpha(r)}{e}~,
\end{equation}
where $n$ is the winding number of the string which is set to one throughout this paper. To construct the loop we use the profile $f(r)$ from the static string\footnote{See Appendix \ref{sect:Static String} for details.}. 
After making the conformal metric ansatz 
\begin{equation}
\gamma_{ij}dx^idx^j= \chi(dx^2+dy^2+dz^2)~,
\end{equation} 
we solve the Hamiltonian constraint to obtain the conformal factor $\chi$.

\section{Results} \label{sect:collapse}

We simulate the collapse of circular loops, scanning through the initial condition parameter radius $R$ and the model symmetry-breaking scale $\eta$ (and hence string tension via \eqn{eqn:Gmu}),
in the critical coupling limit with $e=1$ and $\lambda=2$. The loop begins at rest but quickly accelerates to close to  the speed of light due mainly to the string tension. We find this motion to be consistent with the Nambu-Goto action dynamics (see Appendix \ref{sec:Nambu-Goto})
\begin{equation}\label{eqn:Nambu-goto}
r = R \cos \frac{\tau}{R}~,
\end{equation}
up to $r\sim\delta$ which is the thickness of the string given by
\begin{equation}
\delta = \frac{1}{\eta \sqrt{\lambda}}~,
\end{equation}
and $\tau$ is the time coordinate at spatial infinity. Depending on the choice of $\mu$ and $R$, there are two possible outcomes: $(i)$ the string unwinds itself and the resulting radiation disperses or  $(ii)$ a BH forms. 

\begin{figure}[tb]
\begin{center}
{\includegraphics[width=1\columnwidth]{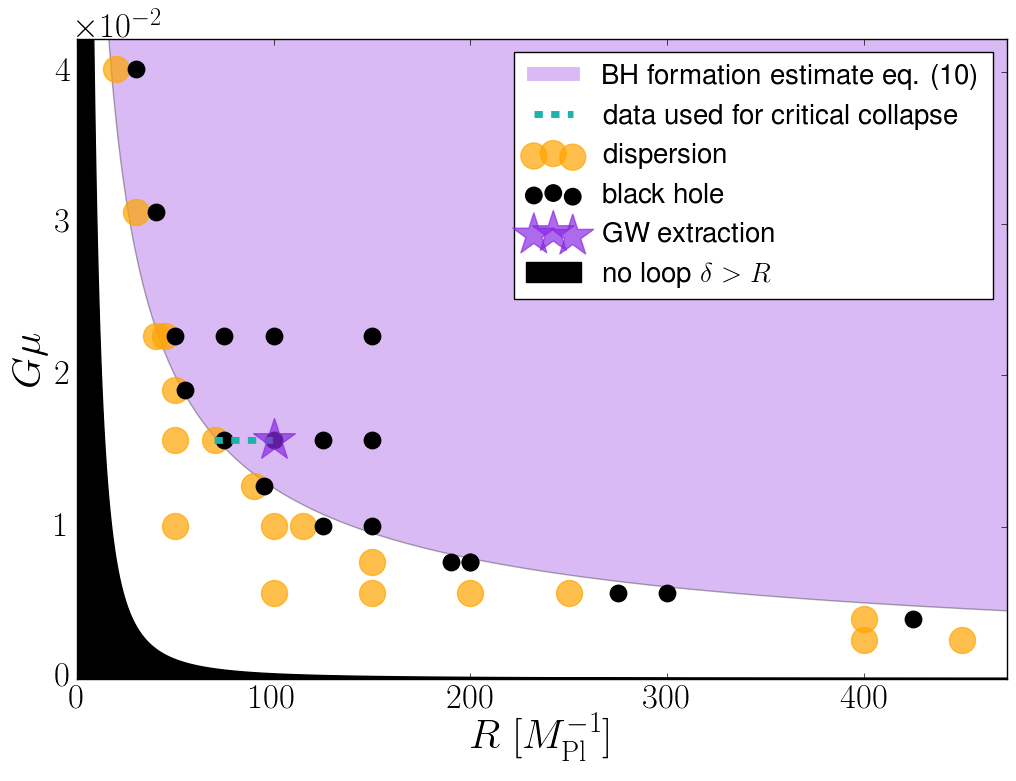}}
\caption{\textbf{Overview of simulations :} The loop can either form a BH or unwind and radiate all its mass. The analytical expression derived from the hoop conjecture accurately predicts the outcome. Movie links for the evolution over time of the collapse  are available for the \underline{\color{blue} \href{https://youtu.be/nHH3gTEjMPo}{dispersion}} \cite{VidDis} and  \underline{\color{blue} \href{https://youtu.be/U5CkThsDU6w}{black hole}} \cite{VidBH} cases.}
 \label{fig:RvsEta}
\end{center}
\end{figure}
This result can be predicted using the hoop conjecture as follows.  A BH forms if the loop mass $M_{\mathrm{\mathrm{loop}}} =2\pi \mu R$ is enclosed within a radius smaller than its Schwarzschild radius $2GM_{\mathrm{loop}}$. In addition, the smallest volume in which a loop can be contained before the string unwinds has radius $\delta$, which sets the Schwarzschild radius the lower bound for BH formation to be $2G M_{\mathrm{\mathrm{loop}}}>\delta$, or 
\begin{equation}\label{eqn:BlackHoleAnalytic}
R > \sqrt{\frac{1}{8 \pi\lambda}}(G\mu)^{-3/2} \mpl^{-1}~.
\end{equation}
Moreover, as the minimum radius of a loop is $R=\delta$, we don't expect dispersion cases for $G\mu>(4\pi)^{-1}$ and all loops will form BHs.
We find this estimate to be a good predictor (see fig. \ref{fig:RvsEta}), which suggests that black hole formation is broadly independent of field interactions.

If a black hole forms,  the amount of initial mass that falls into the black hole depends on the initial radius $R$ for fixed $G\mu$, with the rest being radiated in either gravitational waves or matter.

\begin{figure}[tb]
\begin{center}
{\includegraphics[width=1\columnwidth]{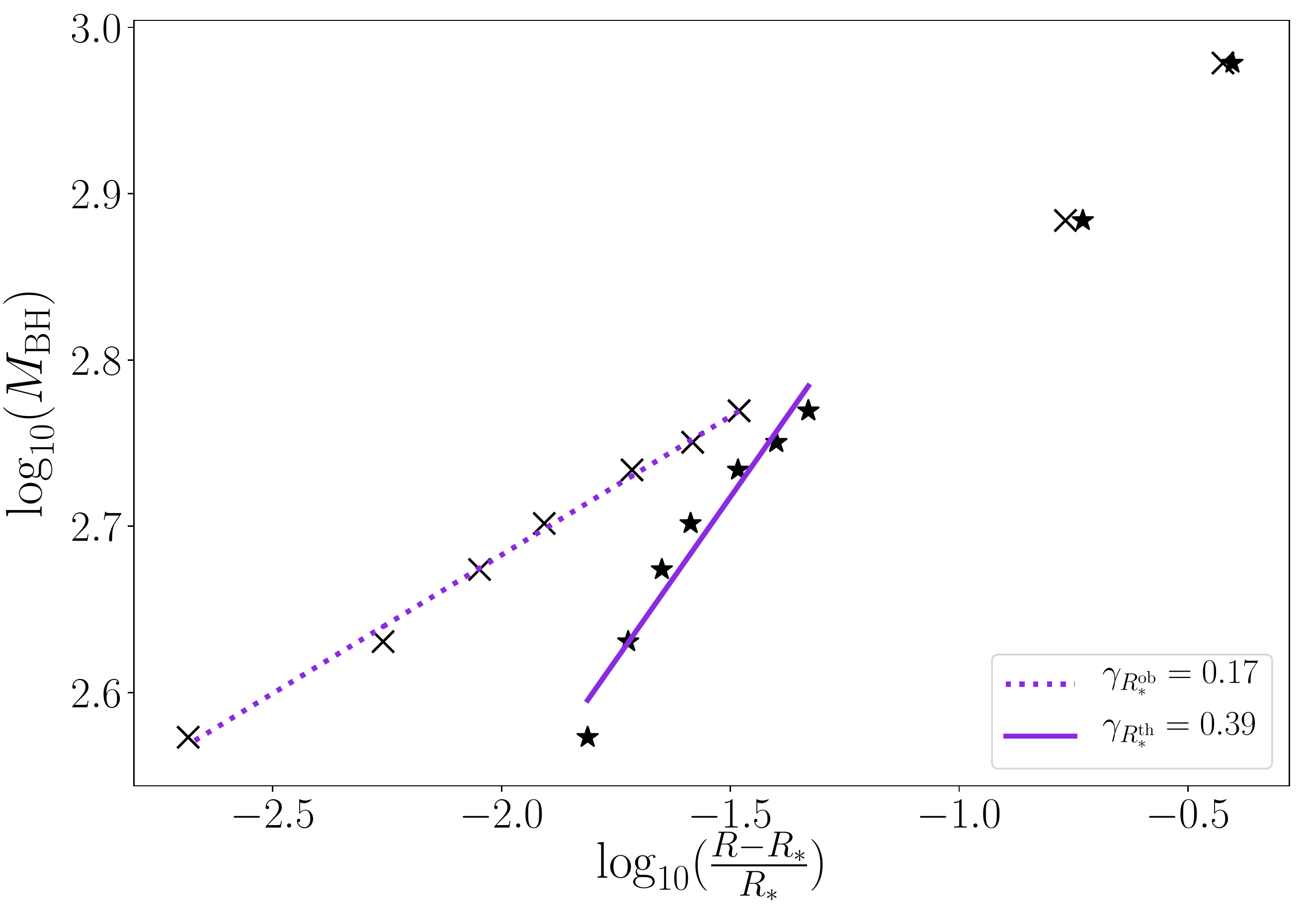}}
  \caption{{\bf Critical collapse}: We plot the logarithm of the mass of the black hole vs the logarithm of the difference between the initial and the theoretical(star)/observed(cross) critical radius for $G\mu=1.6\times 10^{-2}$. As we argued in the text, our simulation showed that the actual $R_*^{\mathrm{ob}} > R_*^{\mathrm{th}}$, resulting in a critical index within $0.17<\gamma<0.39$, where the error is due to the uncertainty in determining numerically $R_*^{\mathrm{th}}<R_*<R_*^{\mathrm{ob}}$. Note that we only use the first 7 points to compute the critical index for $R\leq 0.05R_*$ as the critical relation is only expected to hold perturbatively.}
 \label{fig:Ulrel}
\end{center}
\end{figure}

We investigate whether this collapse is a Type I or Type II transition \cite{Gundlach:1999cu} by studying the mass of the black hole close to the critical radius. Supposing it is a Type II collapse and let $R_*$ be the critical point such that $M_{\mathrm{BH}}(R_*)=0$, one can compute the critical index $\gamma$ defined by 
\begin{equation}
M_{\mathrm{BH}} \propto (R-R_*)^{\gamma}~.
\end{equation}
The value asumming the theoretical prediction of \eqn{eqn:BlackHoleAnalytic}, $R_*^{\mathrm{th}} = \sqrt{1/8\pi\lambda}(G\mu)^{-3/2}\mpl^{-1}$, is $\gamma=0.39$, see fig. \ref{fig:Ulrel}. However, in our simulations we have observed $R_*^{\mathrm{ob}}>R_*^{\mathrm{th}}$, giving $\gamma=0.17$, showing that $\gamma$ is highly dependent on the choice of the actual value of $R_*$ -- of which we are unable to identify with confidence due to the lack of computational resources. Therefore, we conclude that $\gamma=0.28\pm 0.11$.

In the subcritical limit where $2GM_{\mathrm{loop}} < \delta$, the loop unwinds as it collapses, transferring all the mass into matter and gravitational radiation. If $R\gg \delta$ the velocity at unwinding is much larger than the escape velocity and all the energy is radiated away. However, if $R\sim\delta$, the velocity can be small enough so that instead of full dispersal the mass slowly decays at the center and a soliton might form.

\section{Gravitational Waves from black hole formation}

We compute the gravitational waveform from the collapse of a loop with $G\mu= 1.6\times 10^{-2}$ and $R = 100~ \mpl^{-1} $ into a black hole, fig. \ref{fig:WaveForm_sub}. Post formation of the apparent horizon, the waveform exhibits the characteristic quasi-normal mode decay, with the dominant mode being the $l=2,m=0$ mode as usual. We found the integrated energy of the signal to be
\begin{equation}
\epsilon \equiv \frac{E_{\mathrm{GW}}}{M_{\mathrm{loop}}} = 0.5\pm 0.2~\%~. \label{eqn:epsilon}
\end{equation}
The error bars come primarily from the presence of the  spurious modes from the initial data mixing in with the early part of the collapse (grey area in fig. \ref{fig:WaveForm_sub}). Even though the velocity of the loop at collision is ultra-relativistic, $\sim 0.99~c$, the GW production is strongly suppressed when compared to other ultra-relativistic events. For comparison, a boosted head-on black hole merger ($14\pm3\%$) and relativistic fluid particle collapse ($16\pm 2\%$) radiates a much larger fraction of its total mass in gravitational waves \cite{Sperhake:2008ga,East:2012mb}. This suggests that the initial apparent horizon is very spherical -- possibly due to the thickness of our strings when compared to the Schwarzschild radius, i.e. $2GM_{\mathrm{loop}} \sim {\cal O}(1) \times \delta$. In the limit of infinitisimally thin strings, the maximum GW production was calculated by Hawking to be 29\% \cite{HAWKING199036}. Hence, we believe that one can boost the efficiency by colliding thinner strings (i.e. $2GM_{\mathrm{loop}}\gg \delta$) -- in this limit the hoop conjecture argument above suggests that a black hole will form before the loop has a chance to interact and unwind, thus it is possible that the GW emission will be larger  via Hawking's argument, though this has not been demonstrated numerically.

Finally, loops in general are generated non-circularly with many different oscillating stable configurations. Nevertheless, in the presence of gravity, we expect gravity to eventually win out, with roughly the timescale of their gravitational collapse to be the free-fall time-scale. In the final stages of collapse, we expect the tension to circularize the loops and thus our results should hold in general.

\section{Discussion and detection prospects} \label{sect:discussion}

We have extracted the gravitational wave signal for the case $G\mu = 1.6\times 10^{-2}$, and  $R = 100~ \mpl^{-1}$ and found that the efficiency  $\epsilon=0.5\pm 0.2 ~\%$ of the initial mass is radiated into gravitational waves. The QNM frequency of our GW waveform (fig. \ref{fig:WaveForm_sub}) is in the UV range and out of any current or future detectors.  On the other hand, if we assume that our numerical results scale, we can ask whether we can detect suitably massive cosmic strings loops with current or future detectors. The two key parameters are $(i)$ the frequency and $(ii)$ the luminosity of the event, both which depend on the masses. The former constraints our loop parameter space to $2\pi\mu R \approx M_{detector}$. We choose $M_{detector}$ such that its frequency lies at peak sensitivity of LIGO/VIRGO ($f\sim~100Hz$). For the latter,  the strain $h$ observed at a distance $d$ from a source of GWs is
\begin{equation}
\left(\frac{h}{10^{-21}}\right)\sim \sqrt{\frac{E_{\mathrm{GW}}}{3\times 10^{-3} M_\odot}}\left(\frac{10~\mathrm{Mpc}}{d}\right)~. \label{eqn:distance_d}
\end{equation}
Cosmic strings loops are generated during the evolution of the string network when strings intercommute, although there is presently no consensus on the probability distribution of loops and their classification (see e.g. \cite{Blanco-Pillado:2017oxo,Ringeval:2017eww}). Furthermore, it is not clear that all loops will collapse due to the presence of non-intersecting loop configurations and the uncertainty in their angular momentum loss mechanisms. Hence, we will take the agnostic view that only planar loops will collapse -- assuming that planar loops will circularize as argued by \cite{Hawking1989}. Suppose then $N(R,z)$ is the co-moving production density rate of planar loops of radius $R$ at redshift $z$ (i.e. it has dimensions $[N(R,z)] = L^{-3}T^{-1}$), then the detection rate is given by
\begin{equation}
\Gamma = \int_{0}^{z_d}4\pi \left[\int_{0}^{z}\frac{dz'}{H(z')}\right]^2\frac{N(r,z)dz}{H(z)}~,~d= \int_{0}^{z_d}\frac{dz}{(1+z)H}~.
\end{equation}
such that $z_d$ is the maximum range in redshift of the detector, which  itself depends on the energy of the GW $E_{\mathrm{GW}}$ emitted. Our numerical results \eqn{eqn:epsilon} suggest that $0.5\%$ of the total string loop mass is emitted, which is an order of magnitude smaller than that of the typical BH-BH mergers, translating to about a factor of 3 shorter in detectable distance $d$. For LIGO/VIRGO and ET, the maximum redshift range is then $z_d\sim 0.005$ and $z_d\sim 0.05$ respectively. In this limit, $\Gamma$ can be approximated as
\begin{equation}
\Gamma \approx \epsilon^{3/2}\left(\frac{R}{GM_{\odot}}\right)^{3/2}(G\mu)^{3/2}\left(\frac{10^{-19}}{h}\right)^{3}\left(\frac{N(R,z)}{\mathrm{Mpc}^{-3}}\right)~.
\end{equation}
Clearly, $\Gamma$ depends linearly on $N(R,z)$, which itself depends on the cosmic string model and its network evolution, which at present is still being debated vigorously as mentioned above. For example, in \cite{Hawking1989}, it was estimated that $N(R,z)\propto (G\mu)^{2R/s-4}$ where $s$ is the correlation length of the loop. Other estimates are given in \cite{Polnarev1991, Caldwell1996}.  On the other hand, we can use the non-detection of such collapse events in the present LIGO/VIRGO to put a constraint on $N(R,z)$. For $G\mu\sim 10^{-10}$ which leads to solar system sized loops of $R\sim {\cal O}(100)$ a.u., this is $N(R,z)< 10^{-2}~ \mathrm{Gpc}^{-3}~\mathrm{yr}^{-1}$, which is a lower detection rate than what is expected from BH mergers of ${\cal O}(10)~ \mathrm{Gpc}^{-3}~\mathrm{yr}^{-1}$ \cite{LIGOScientific:2018mvr}.

Finally, we note that this is a conservative estimate since these solar system sized loops satisfy $R_{\mathrm{BH}}\sim{\cal O}(10^{40}) \times \delta$ and hence are thin loops.  In this limit, $\epsilon$ might be closer to $29~\%$, with a corresponding increase in $d$.  We will numerically investigate the collapse of these thin loops in a future work.\\

\acknowledgments
We acknowledge useful conversations with Jos\'e Juan Blanco-Pillado, Katy Clough, Ed Copeland, William East, Mark Hindmarsh, Paul Shellard, James Widdicombe and Helvi Witek. EAL is supported by an STFC AGP-AT grant (ST/P000606/1).
We would also like to thank the GRChombo team
\href{http://www.grchombo.org}{(http://www.grchombo.org/)} and the
COSMOS team at DAMTP, Cambridge University for
their ongoing technical support.  Special thanks to Federica Albertini, Gregorio Carullo and Alastair Wickens. Numerical simulations
were performed on the COSMOS supercomputer and the Cambridge CSD3 HPC, funded by DIRAC3/BIS,  on BSC Marenostrum IV via PRACE grant Tier-0 PPFPWG, by the Supercomputing Centre of Galicia and La Palma Astrophysics Centre via BSC/RES grants AECT-2017-2-0011 and AECT-2017-3-0009 and  on SurfSara Cartesius under Tier-1 PRACE grant DECI-14 14DECI0017. 

\appendix
\section{Numerical Methodology}

\subsection{Evolution Equations}\label{EMevolutionEquations}
In this work, we use \textsc{GRChombo}, a multipurpose numerical relavity code \cite{Clough:2015sqa} which solves the BSSN \cite{Baumgarte:1998te,PhysRevD.52.5428,Shibata:1995we} formulation of the Einstein equation. The 4 dimensional spacetime metric is decomposed into a spatial metric on a 3 dimensional spatial hypersurface, $\gamma_{ij}$, and an extrinsic curvature $K_{ij}$, which are both evolved along a chosen local time coordinate $t$. The line element of the decomposition is
\begin{equation}
ds^2=-\alpha^2\,dt^2+\gamma_{ij}(dx^i + \beta^i\,dt)(dx^j + \beta^j\,dt)~,
\end{equation}
where $\alpha$ and $\beta^i$ are the lapse and shift, gauge parameters.  These gauge parameters are specified on the initial hypersurface and then allowed to evolve using gauge-driver equations, in accordance with the puncture gauge \cite{Campanelli:2005dd,Baker:2005vv}, for which the evolution equations are
\begin{align} \label{eqn:MovingPuncture}
&\partial_t \alpha = - \mu \alpha K + \beta^i \partial_i \alpha ~ , \\
&\partial_t \beta^i = B^i ~ , \\
&\partial_t B^i = \frac{3}{4} \partial_t \Gamma^i - \eta B^i ~ ,
\end{align}
where the constants $\eta$ and $\mu$ are of order $1/M_{\mathrm{ADM}}$ and unity respectively.

The induced metric is decomposed as 
\begin{equation}
\gamma_{ij}=\frac{1}{\chi}\tilde\gamma_{ij}~,~\det\tilde\gamma_{ij}=1~,~ \chi = \left(\det\gamma_{ij}\right)^{-\frac{1}{3}}  ~ .
\end{equation}
The extrinsic curvature is decomposed into its trace, $K=\gamma^{ij}\,K_{ij}$, and its traceless part $\tilde\gamma^{ij}\,\tilde A_{ij}=0$ as
\begin{equation}
K_{ij}=\frac{1}{\chi}\left(\tilde A_{ij} + \frac{1}{3}\,K\,\tilde\gamma_{ij}\right) ~ .
\end{equation}
The conformal connections are $\tilde\Gamma^i=\tilde\gamma^{jk}\,\tilde\Gamma^i_{~jk}$ where $\tilde\Gamma^i_{~jk}$ are the Christoffel symbols associated with the conformal metric $\tilde\gamma_{ij}$.
The evolution equations for the gravity sector of BSSN are then
\begin{align}
&\partial_t\chi=\frac{2}{3}\chi \alpha K-\frac{2}{3}\chi \partial_k\beta^k+\beta^k\partial_k\chi ~ , \label{eqn:dtchi2} \\
&\partial_t\tilde\gamma_{ij} =-2\,\alpha\, \tilde A_{ij}+\tilde \gamma_{ik}\,\partial_j\beta^k+\tilde \gamma_{jk}\,\partial_i\beta^k \nonumber \\
&\hspace{1.3cm} -\frac{2}{3}\,\tilde \gamma_{ij}\,\partial_k\beta^k +\beta^k\,\partial_k \tilde \gamma_{ij} ~ , \label{eqn:dttgamma2} \\
&\partial_t K = -\gamma^{ij}D_i D_j \alpha + \alpha\left(\tilde{A}_{ij} \tilde{A}^{ij} + \frac{1}{3} K^2 \right) \nonumber \\
&\hspace{1.3cm} + \beta^i\partial_iK + 4\pi\,\alpha(\rho + S) \label{eqn:dtK2} ~ , \\
&\partial_t\tilde A_{ij} = \chi\left[-D_iD_j \alpha + \alpha\left( R_{ij} - 8\pi\,\alpha \,S_{ij}\right)\right]^\textrm{TF} \nonumber \\
&\hspace{1.3cm} + \alpha (K \tilde A_{ij} - 2 \tilde A_{il}\,\tilde A^l{}_j)  \nonumber \\
&\hspace{1.3cm} + \tilde A_{ik}\, \partial_j\beta^k + \tilde A_{jk}\,\partial_i\beta^k \nonumber \\
&\hspace{1.3cm} -\frac{2}{3}\,\tilde A_{ij}\,\partial_k\beta^k+\beta^k\,\partial_k \tilde A_{ij}\,   \label{eqn:dtAij2} ~, \\ 
&\partial_t \tilde \Gamma^i=2\,\alpha\left(\tilde\Gamma^i_{jk}\,\tilde A^{jk}-\frac{2}{3}\,\tilde\gamma^{ij}\partial_j K - \frac{3}{2}\,\tilde A^{ij}\frac{\partial_j \chi}{\chi}\right) \nonumber \\
&\hspace{1.3cm} -2\,\tilde A^{ij}\,\partial_j \alpha +\beta^k\partial_k \tilde\Gamma^{i} \nonumber \\
&\hspace{1.3cm} +\tilde\gamma^{jk}\partial_j\partial_k \beta^i +\frac{1}{3}\,\tilde\gamma^{ij}\partial_j \partial_k\beta^k \nonumber \\
&\hspace{1.3cm} + \frac{2}{3}\,\tilde\Gamma^i\,\partial_k \beta^k -\tilde\Gamma^k\partial_k \beta^i - 16\pi\,\alpha\,\tilde\gamma^{ij}\,S_j ~ . \label{eqn:dtgamma2}
\end{align} 

\begin{figure}[t!]
\begin{center}
{\includegraphics[width=1.0\columnwidth]{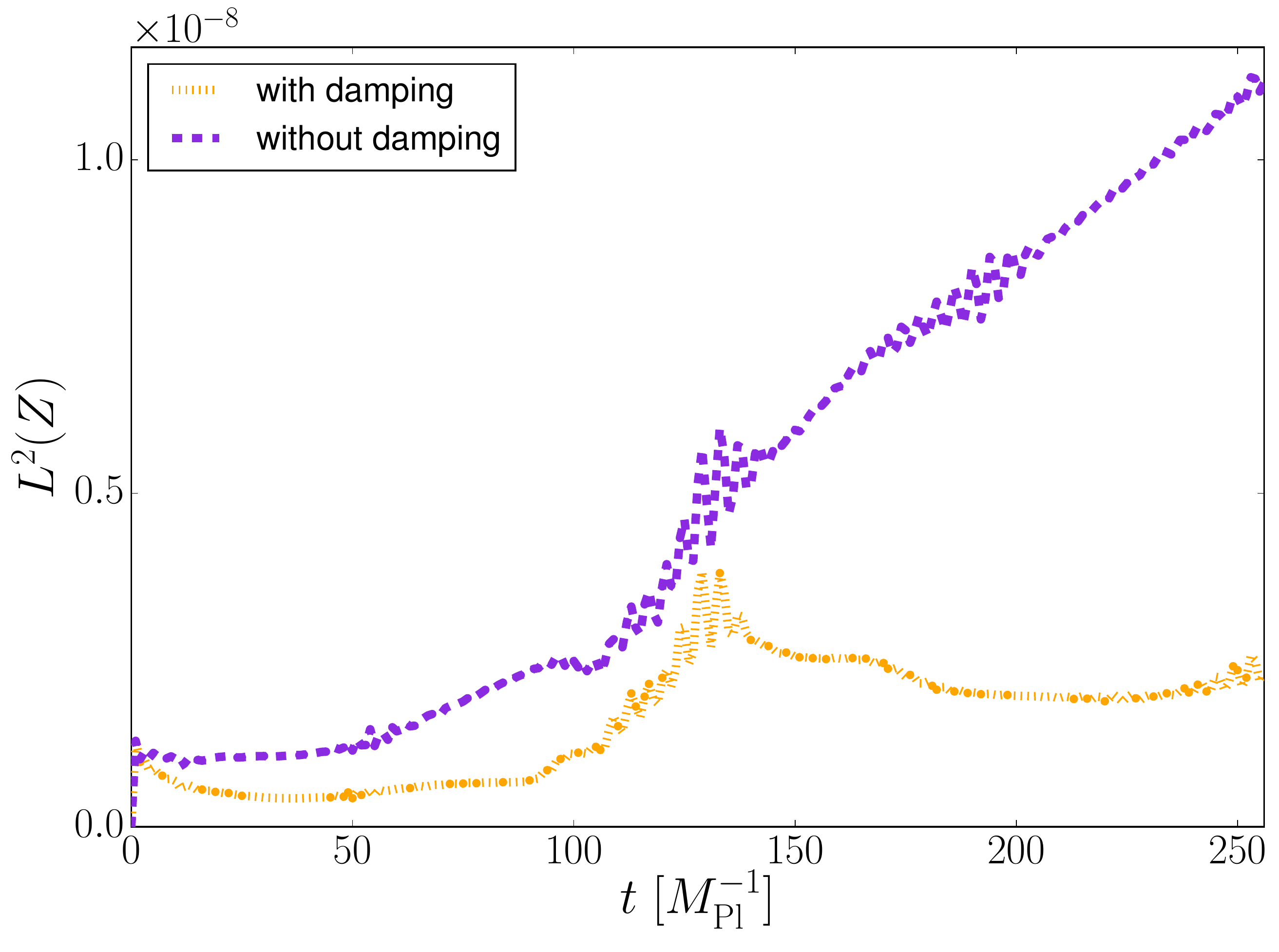}}
\vspace{-1.5em} \caption{\textbf{Gauss constraint for static string:} We run the same simulation for am infinite static string with $G\mu = 1.6\times 10^{-2}$ ($\eta=0.05\mpl$) with and without damping. We find that the damping stabilises the linear growth in violation. }
\label{fig:Zvec_Straigt_String}
\end{center}
\end{figure}
Meanwhile, the matter part of the Lagrangian is 
\begin{equation}
\mathcal{L}_m=-(D_{\mu}{\phi})^{*}(D^{\mu}{\phi})-\frac{1}{4}F_{\mu\nu}F^{\mu\nu}-V(\phi)~,
\end{equation}
which gives the evolution equations 
\begin{equation}
-D_\mu D^\mu\phi+\frac{\partial V(\phi)}{\partial \bar{\phi}}=0~,
\end{equation}
\begin{equation}\label{equ:ElDym}
\nabla_{\mu}F^{\mu\nu} = - e J^{\nu}~,
\end{equation}
with 
\begin{equation}
J^{\nu} = 2Im(\phi^{*}D^{\nu}\phi) ~, ~ F_{\mu\nu} = \partial_{\mu}A_{\nu}-\partial_{\nu}A_{\mu}~.
\end{equation}
We decompose these equations in 3+1 coordinates, following \cite{Zilhao:2015tya}. Furthermore, we impose the Lorenz condition
\begin{equation}
\nabla^{\mu}A_{\mu} = 0~.
\end{equation}
Using the projector 
\begin{equation}
P_{\mu}^{\nu} = \delta_{\mu}^{\nu}+n_{\mu}n^{\nu}~,
\end{equation}
where $n^{\mu}$ is the normal to the hypersurface, the gauge field and current can further be decomposed into traverse and longitudinal components via
\begin{equation}
\begin{split}
A_{\mu} &= \mathcal{A}_{\mu}+n_{\mu} \mathcal{A} ~,\\
J_{\mu} &= \mathcal{J}_{\mu}+n_{\mu} \mathcal{J} ~,\\
\end{split}
\end{equation}
such that
\begin{equation}
\begin{split}
\mathcal{A}_{\mu} &= P_{\mu}^{\nu} A_{\nu}\quad \text{and} \quad \mathcal{A} = -n^{\nu} A_{\nu}~, \\
\mathcal{J}_{\mu} &= P_{\mu}^{\nu} J_{\nu}\quad \text{and} \quad \mathcal{J} = -n^{\nu} J_{\nu}~. 
\end{split}
\end{equation}

The electric and magnetic fields are defined as
\begin{equation}
E_{\mu}= P_{\mu}^{\nu}n^{\rho}F_{\nu\rho}~ ,
\end{equation}
\begin{equation}
B_{\mu} = P_{\mu}^{\nu}n^{\rho}(\star F_{\nu\rho})~ ,
\end{equation}
where $(\star F_{\nu\rho})$ is the dual Maxwell tensor. Using the previous decomposition we rewrite the Maxwell tensor as 
\begin{equation}
F_{\mu\nu} = n_{\mu}E_{\nu}-n_{\nu}E_{\mu}+\partial_{\mu}\mathcal{A}_{\nu}
-\partial_{\mu}\mathcal{A}_{\nu}~.
\end{equation}
In addition, \eqn{equ:ElDym} gives the Gauss constraint 
\begin{equation} 
\tilde{\nabla}_i E^i = e\mathcal{J}~, \label{eqn:gauss_con}
\end{equation}
where $ \tilde{\nabla} = P_{\mu}^\nu\nabla_\nu$. 

To ensure that numerical violation of \eqn{eqn:gauss_con} is kept to a minimum, we stabilise it by introducing an auxiliary damping variable $Z$ \cite{Zilhao:2015tya,Hilditch:2013sba,Palenzuela:2009hx}, resulting in the following modified evolution equations
\begin{align}
&\partial_t E^i = \alpha(E_i-e\mathcal{J}^i+\tilde{\nabla}_i\mathcal{A})-\mathcal{A}\tilde{\nabla}_i\alpha
+ \beta^j\partial_j E^i\nn
&\hspace{1.3cm} - E^j \partial_j\beta^i~,\\
&\p_t\mathcal{A} = -\mathcal{A}^i\tilde{\nabla}_i\alpha + \alpha (K\mathcal{A}-\tilde{\nabla}_i\mathcal{A}^i-Z)+\beta^j\partial_j\mathcal{A}~,\\ 
&\p_t\mathcal{A}_i = -\alpha(E_i+\tilde{\nabla}_i\mathcal{A})-
\mathcal{A}\tilde{\nabla}_i\alpha+\beta^j\partial_j\mathcal{A}_i\nn
&\hspace{1.3cm}+\partial_i\beta^j\mathcal{A}_j~, \\ 
&\p_t Z = \alpha (\tilde{\nabla}_i E^i-e\mathcal{J}-\kappa Z )+\beta^j\partial_jZ~.
\end{align} 
From fig. \ref{fig:Zvec_Straigt_String} we see the scheme is effective at stopping the growth of constraint violations.

Finally, we decompose the complex scalar field  
\begin{equation}
\phi = \frac{1}{\sqrt{2}}\left(\phi_1+i\phi_2\right)~,
\end{equation}
and rewriting the matter equation with BSSN variables, 

\begin{figure}[tb]
\begin{center}
{\includegraphics[width=1.0\columnwidth]{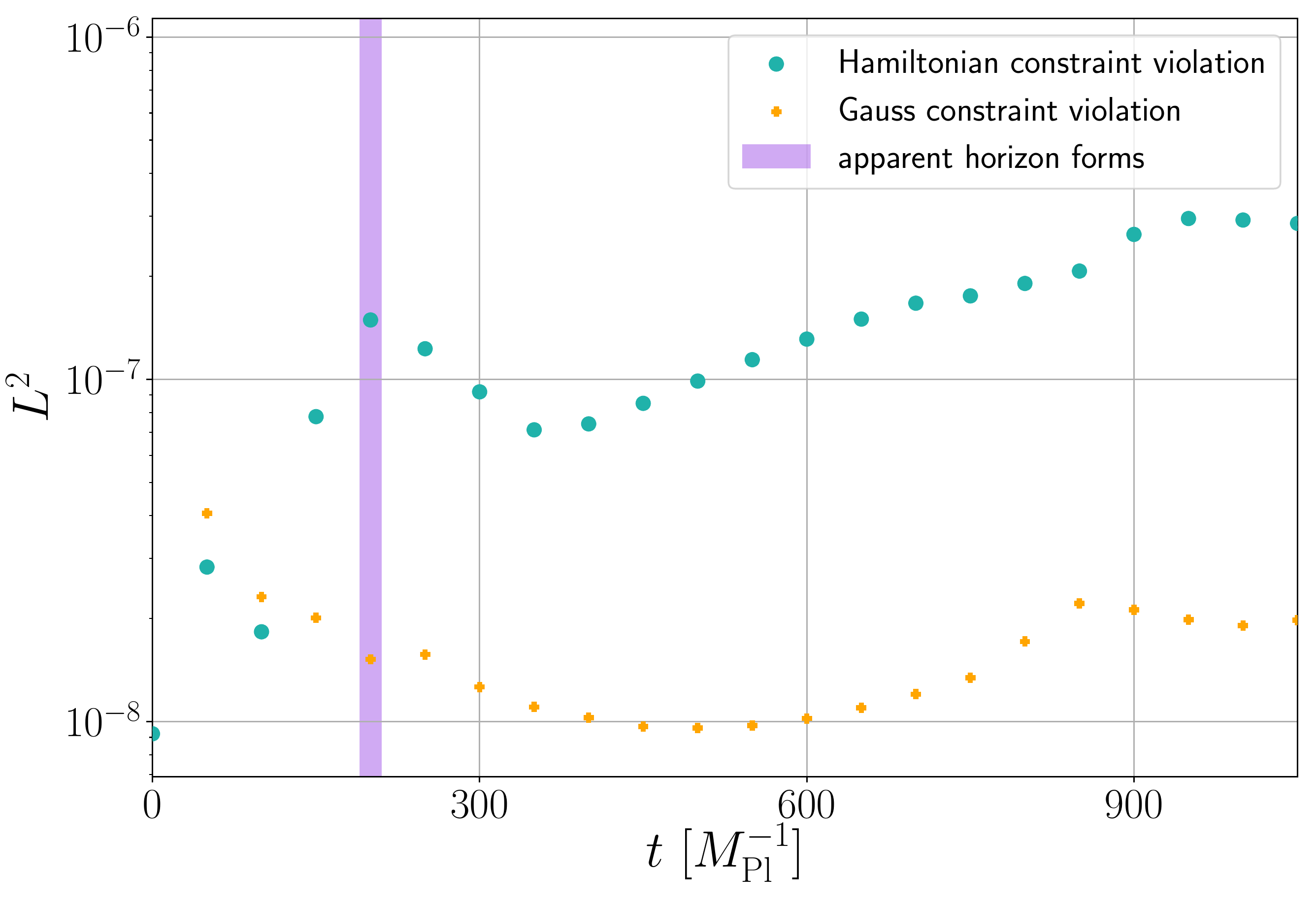}}
\vspace{-1.5em} \caption{\textbf{$L^2$ norm of constraints:} Loop with $G\mu=1.6\times 10^{-2}$ and $R=100~\mpl^{-1}$ remains stable throughout evolution, even after black hole formation. The initial Hamiltonian constraint is smaller than it can be maintained by the evolution scheme. The momentum constraints violation are negligible throughout.}
\label{fig:L2H}
\end{center}
\end{figure}
\begin{align}
&\partial_t \phi_a = \alpha \Pi_{M,a} +\beta^i\partial_i \phi_a \label{eqn:dtphi2} ~ , \\
&\partial_t \Pi_{M,a}=\beta^i\partial_i \Pi_{M,a} + \alpha\partial_i\partial^i \phi_a + \partial_i \phi_a\partial^i \alpha \nn
&\hspace{1.3cm} +\alpha\left(K\Pi_{M,a}-\gamma^{ij}\Gamma^k_{ij}\partial_k \phi_a+\frac{dV}{d\phi_a}\right) \nn
&\hspace{1.3cm}
+\alpha \large(-e^2A_\mu A^\mu \phi_a \pm e\phi_{a+1}\nabla_\mu A^\mu \nn
&\hspace{1.3cm} \pm 2eA^\mu \partial_\mu\phi_{a+1}\large) \label{eqn:dtphiM2}~,\\
&\partial_t E^i = \alpha K E^i+e\alpha \chi  \tilde{\gamma}^{ij}\mathcal{J}_j+\alpha\chi\tilde{\gamma}^{ij}\partial_jZ
\nn
&\hspace{1.3cm} +\chi^2\tilde{\gamma}^{ij}\tilde{\gamma}^{kl}
\partial_l\alpha(\partial_j\mathcal{A}_k-\partial_k\mathcal{A}_j)  \nn
&\hspace{1.3cm} +\chi^2\tilde{\gamma}^{ij}\tilde{\gamma}^{kl}
(\tilde{D}_k\partial_j\mathcal{A}_l-\tilde{D}_k\partial_l\mathcal{A}_j)  \nn
&\hspace{1.3cm} +\frac{\alpha}{2}\chi\tilde{\gamma}^{ij}\tilde{\gamma}^{kl}(\partial_j\mathcal{A}_l
\partial_k\chi-\partial_k\mathcal{A}_j\partial_l\chi) \nn
&\hspace{1.3cm} + \beta^j\partial_j E^i - E^j \partial_j\beta^i - \alpha e \mathcal{J}^{i}~,\\
&\p_t\mathcal{A} = \alpha K \mathcal{A} -\alpha \chi \tilde{\gamma}^{ij} \partial_j\mathcal{A}_i+\alpha\chi\mathcal{A}_i\tilde{\Gamma}^i-\alpha Z
\nn
&\hspace{1.3cm}+\frac{\alpha}{2}\mathcal{A}_i\tilde{\gamma}^{ij}\partial_j\chi-\chi\tilde{\gamma}^{ij}
\mathcal{A}_i\partial_j\alpha +\beta^j\partial_j\mathcal{A}~, \\
&\p_t\mathcal{A}_i = -\alpha\chi^{-1}\tilde{\gamma}_{ij}E^j- \alpha\partial_i\mathcal{A}-\mathcal{A}\partial_i\alpha \nn
&\hspace{1.3cm}+\beta^j\partial_j\mathcal{A}_i +\partial_i\beta^j\mathcal{A}_j~, \\ 
&\p_t Z = \alpha \tilde{\nabla}_i  E^i -\frac{3}{2}\frac{\alpha}{\chi} E^i \partial_i \chi -\alpha e\mathcal{J}-\alpha\kappa Z+\beta^j\partial_jZ ~,
\end{align} 
where  $a \in \{1,2\}$ and the second order Klein Gordon equation has been decomposed into two first order equations as usual. The stress energy tensor for Abelian Higgs is
\begin{equation}
T_{\mu\nu} = 
D_{(\mu}\phi^{*}D_{\nu)}\phi+F_{\mu\alpha}F_\nu^{\alpha}+g_{\mu\nu}\mathcal{L}_m,
\end{equation}
and its various components are defined as
\begin{align}
&\rho = n_a\,n_b\,T^{ab}\,,\quad S_i = -\gamma_{ia}\,n_b\,T^{ab}\,, \nonumber \\
&S_{ij} = \gamma_{ia}\,\gamma_{jb}\,T^{ab}\,,\quad S = \gamma^{ij}\,S_{ij} ~.
\label{eq:Mattereqns}
\end{align}
The Hamiltonian constraint 
\begin{equation}\label{eq:HamiltonianContraint}
\mathcal{H} = R + K^2-K_{ij}K^{ij}-16\pi \rho ~ ,
\end{equation}
the momentum constraint 
\begin{equation}\label{eq:MomentumContraint}
\mathcal{M}_i = D^j (\gamma_{ij} K - K_{ij}) - 8\pi S_i ~,
\end{equation}
and the Gauss constraint 
\begin{equation}
\mathcal{Z}=\tilde{\nabla}_i E^i+e\mathcal{J}^{\nu}n_{\nu}~,
\end{equation}
are monitored throughout the evolution to check the quality of our simulations (see fig. \ref{fig:L2H}). 
Our boundary conditions are Dirichlet.

 \begin{figure}[tb]
\begin{center}
{\includegraphics[width=1.0\columnwidth]{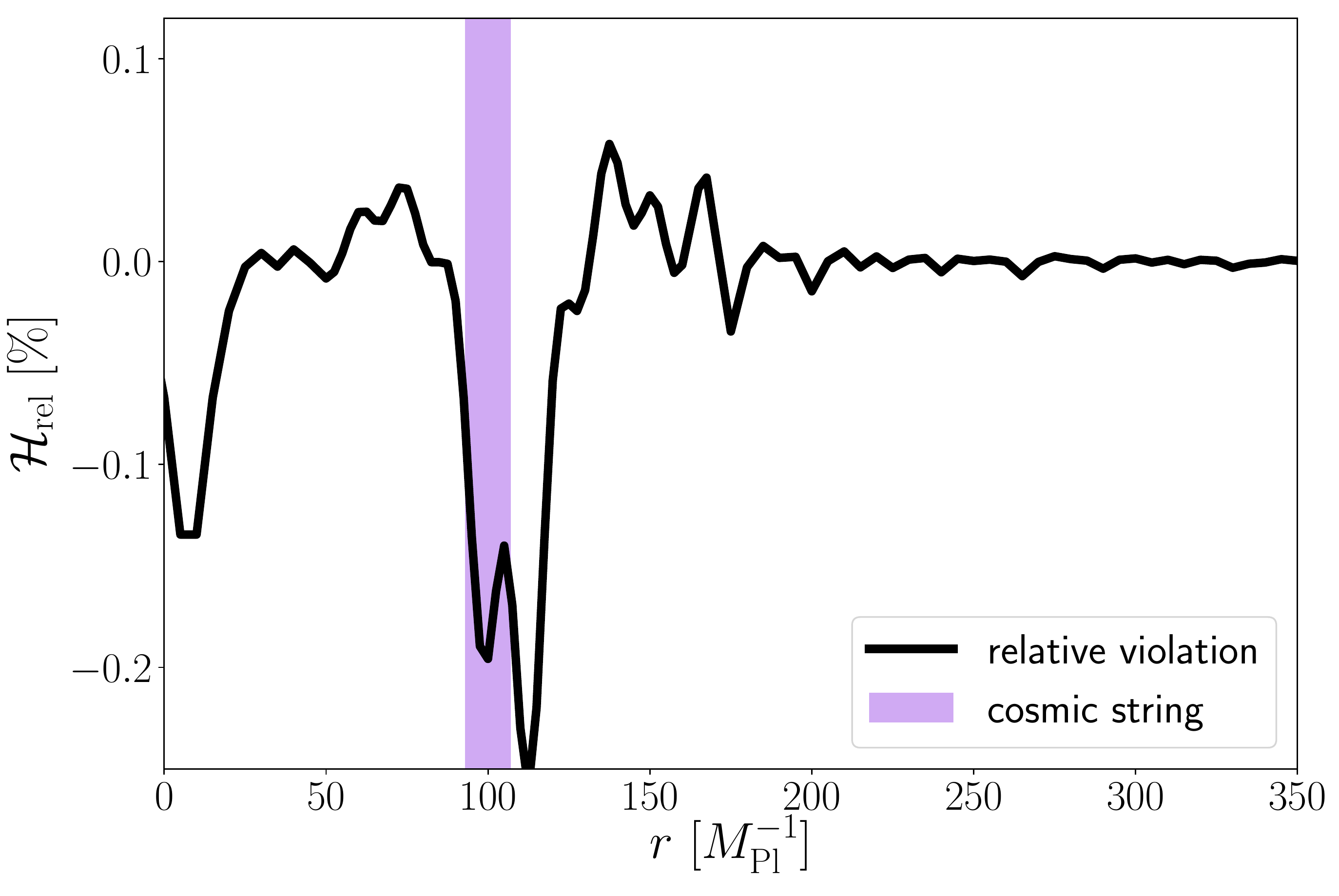}}
\vspace{-1.5em} \caption{\textbf{Initial relative violation}: 
Slice through initial data for loop from center through string with $G\mu$ = $1.6\times 10^{-2}$ and initial radius $R = 100~\mpl^{-1}$. The green region indicates where the string is located. We find that there is an error of at most $0.3 \%$.}
\label{fig:InitialDataHam}
\end{center}
 \end{figure}
\subsection{Initial Data}

\begin{figure}[tb]
\begin{center}
{\includegraphics[width=0.9\columnwidth]{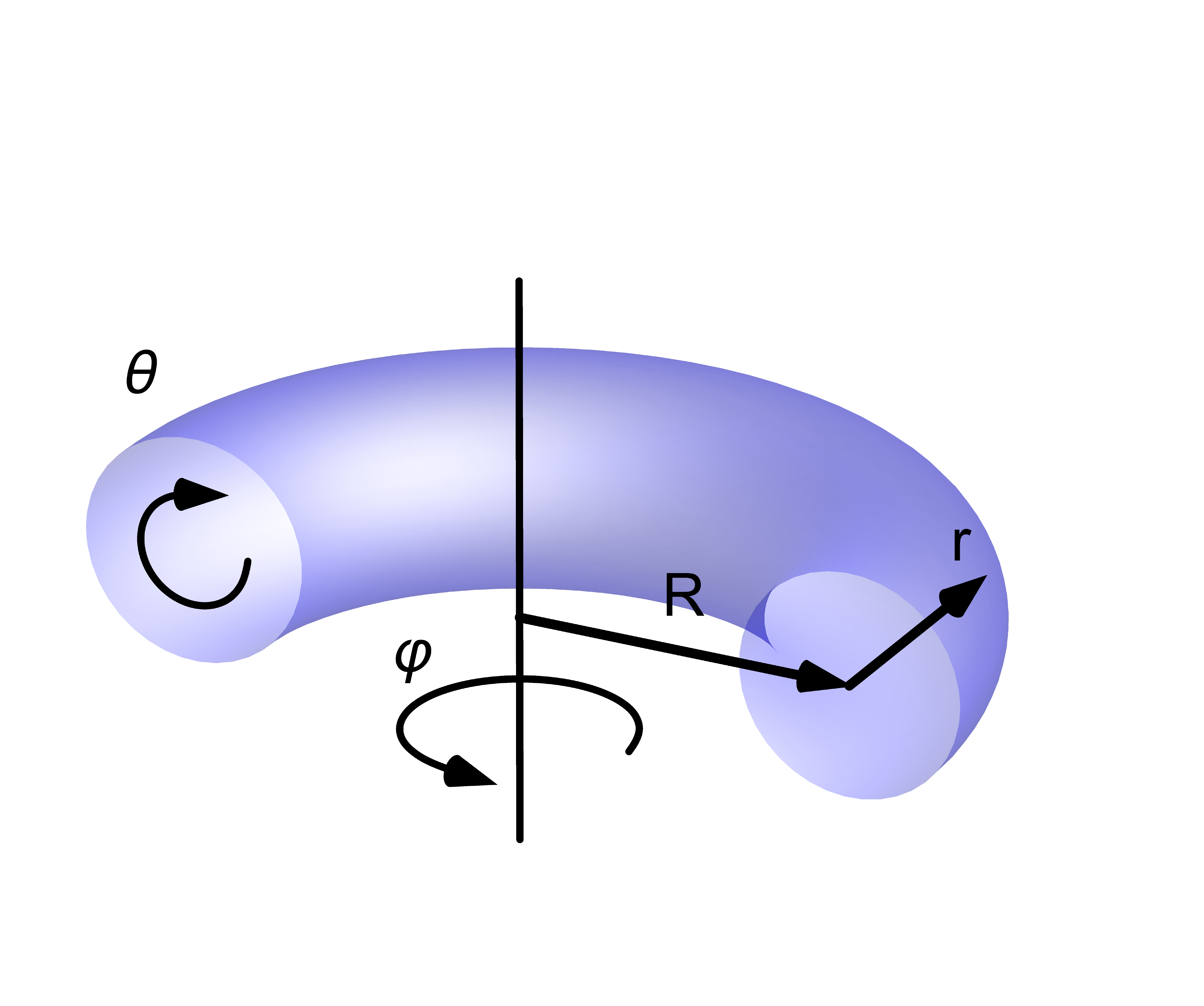}}
\caption{\textbf{Toroidal coordinates} encode the symmetry of our cosmic string loops. They are used to generate the initial field configuration, where R defines the radius of the loop.} 
 \label{fig:Inital_data}
\end{center}
\end{figure}

We set up the field as mentioned in the main text using toroidal coordinates (see fig. \ref{fig:Inital_data}).  Time symmetry is assumed for our initial data,
\begin{equation}
K=0~, \qquad A_{ij}=0~,
\end{equation}
which automatically fulfils the momentum contraint (eq. \ref{eq:MomentumContraint}). In addition, we make a conformally flat\footnote{This is not the unique solution to the constraint equations given the initial field configuration. However, it is the most easily implemented, as more general initial conditions require much greater computational resources to find. Conformal flatness is also consistent with the fact that the spacetime is asymptotically Schwarzschild. }
ansatz  $\tilde{\gamma}_{ij}$,
\begin{equation}
\tilde{\gamma}_{ij} = \delta_{ij}~,
\end{equation}
and  impose  the metric to be identity in the center of the string, similar as the static string (see eq. \ref{eqn:Flat_metric}). We find that doing so reduces possible excitations of the string. For the gravitational wave extraction, we impose the condition  
\begin{equation}
\lim_{r\to\infty} \chi = 1~.
\end{equation}
\begin{figure}[tb]
\begin{center}
{\includegraphics[width=1.0\columnwidth]{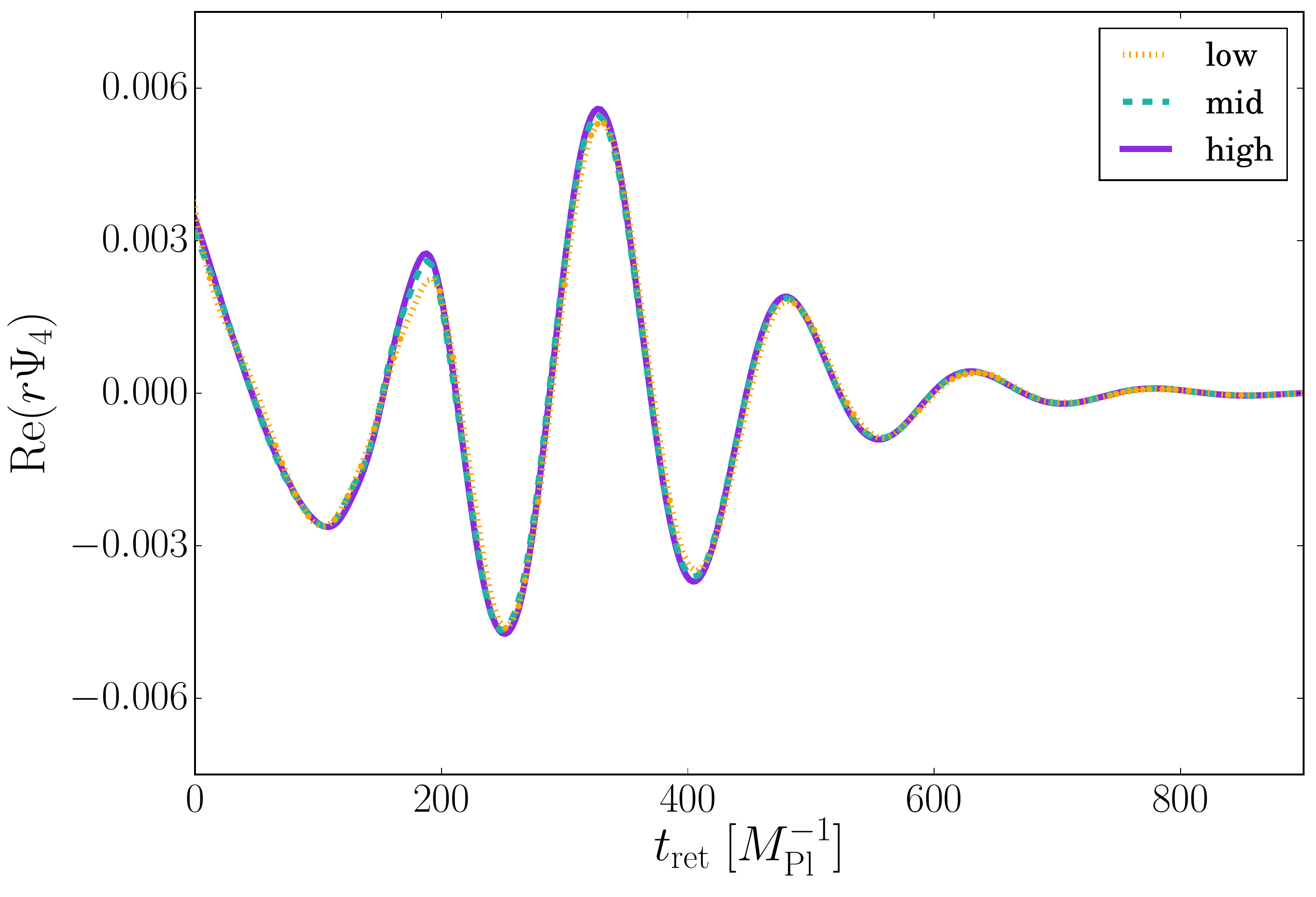}}
\vspace{-1.5em} \caption{\textbf{Convergence in $r\Psi_4$} between low, mid and high resolutions giving an overall 2nd-3rd order convergence. The x-axis $t_{\mathrm{ret}}=t-r_\mathrm{ext}$ is the retarded time where $r_\mathrm{ext}$ is the extraction radius.}
\label{fig:Connvergence}
\end{center}
\end{figure}

We solve for $\chi$ using the Hamiltonian constraint \eqn{eq:HamiltonianContraint}. We reduce the spatial dimension of the problem by using its cylindrical symmetry. This solution is then further relaxed to obtain the final solution, which is that of an excited cosmic string loop.

As shown in fig. \ref{fig:InitialDataHam}, the relative Hamiltonian violation from our prescription is 
\[
\mathcal{H}_{\mathrm{rel}} = \frac{\mathcal{H}}{16 \pi \rho_{\mathrm{max}}}<1\% ~.
\]  

\subsection{Numerical Extraction of Signal}

We extract the Penrose scalar $\Psi_4$ with tetrads proposed by \cite{Baker:2001sf}. Similarly as in black hole binaries, there is some non-physical radiation associated with the initial data, which in our case consists of a toroidal shell of artificial radiation resulting in two GW peaks before the physical signal. While such stray-GW can often be ignored as they quickly radiate away at light speed, in our case due to the rapid collapse of our loops at  ultrarelativistic speeds, they cannot be ignored.

The first peak at $t_{\mathrm{ret}}<0$ is due to this initial radiation travelling opposite to the collapse and could be separated by increasing the loop radius so that the real signal takes longer. However, the second peak (first peak in fig. 1) results from the radiation which travels together with the collapsing loop at similar velocity, which always mixes with the real signal. In any case, increasing the loop radius would result in a cleaner signal but this is computationally very expensive.

To estimate the GW energy we use the equation 
\begin{equation}
\frac{dE_{\mathrm{GW}}}{dt} = \frac{r^2}{16\pi}\int_{\mathcal{S}_r}\left|\int^t_{t_0}\Psi_4dt^{\prime}\right|^2 d\Omega~,
\end{equation}
where $\mathcal{S}_r$ is a sphere of radius $r$.

In the cases for which the cosmic string loop does not form a black hole, most of the matter will escape, typically at velocities close to the speed of light. This scalar and vector radiation overlaps the gravitational wave signal and due to its large mass might leave an imprint on $r\Psi_4$, making the signal extraction problematic\footnote{This could be prevented by setting the extraction zone further out, but this is numerically too expensive.}.

\subsection{Numerics and Convergence Tests}

In fig. \ref{fig:L2H}, we show that the  volume-averaged Hamiltonian constraint violation
\begin{equation}\label{eq:L2H}
L^2(H) =\sqrt{\frac{1}{V} \int_V |\mathcal{H}^2| dV}~,
\end{equation}
where $V$ is the box volume with the interior of the apparent horizon excised, is under control throughout the simulation. 

We use the gradient conditions on $\phi$ and $\chi$ to tag cells for regridding. The precise criteria is chosen depending on the symmetry breaking scale $\eta$ and the total mass of the system. The major distinction for the amount of resolution needed is whether GW are being extracted or not. To obtain a clean GW large boxes are needed to avoid the detection of reflections of the non-physical signal with the boundaries, which increases the cost of the simulation. We double checked that our signal in fig. 1 was consistent with  a $l=2$ $m=0$ QNM \cite{Kokkotas1999} within numerical error. 
\begin{figure}[tb]
\begin{center}
{\includegraphics[width=1.0\columnwidth]{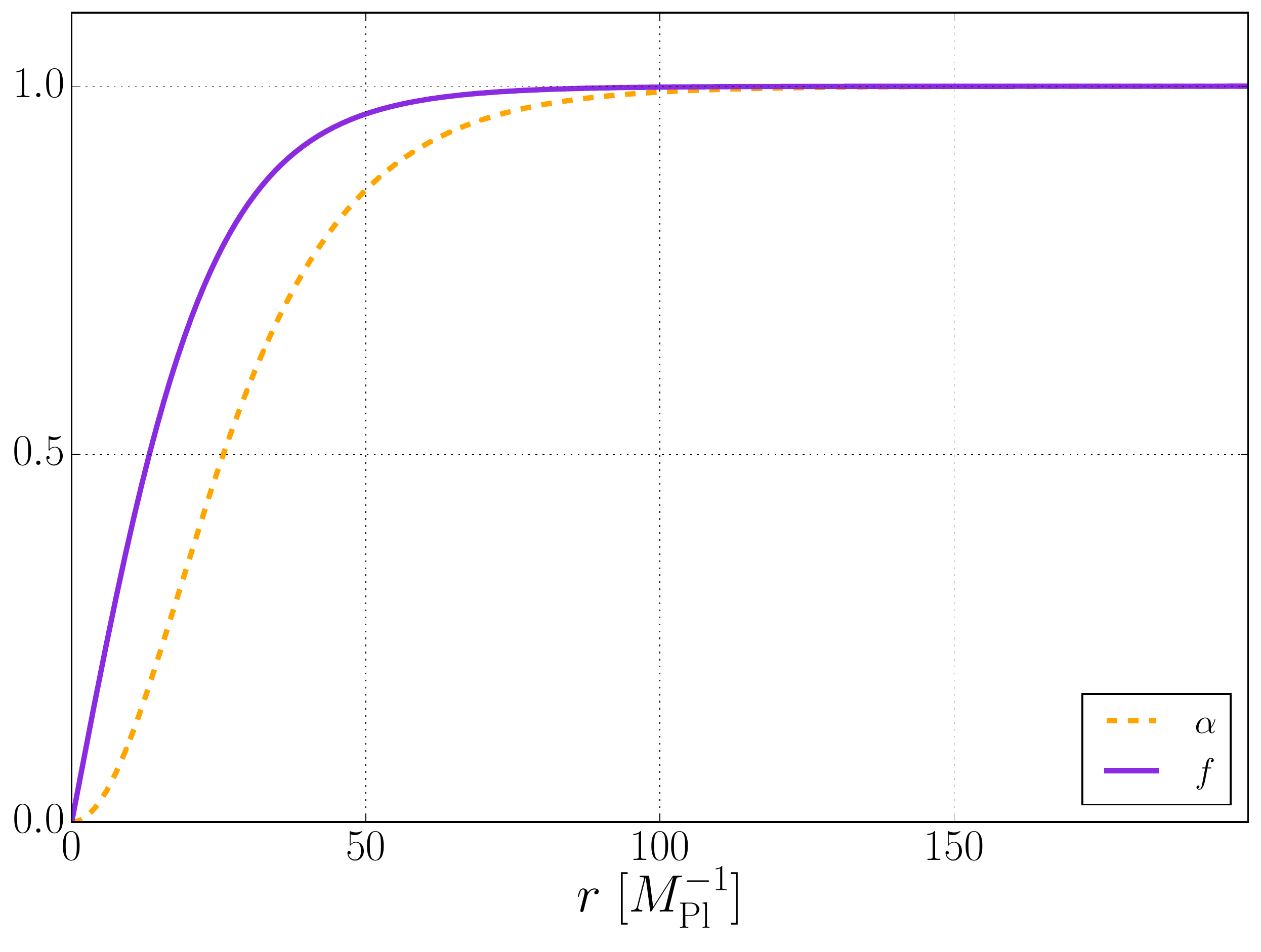}}
\vspace{-1.5em} \caption{\textbf{Radial profile of $\alpha$ and $f$} for an infinite static string with gravity in the critical coupling limit ($e=1,~ \lambda=2$) and $\eta=0.05~\mpl$  ($G\mu=1.6\times 10^{-2}$).}
\label{fig:FAProfile}
\end{center}
 \end{figure}
  
We tested the convergence of our simulations with a cosmic string loop of $\eta = 0.05$ ($G\mu=1.6\times 10^{-2}$) and $R=100~\mpl^{-1}$ by using a box of size $L=2048~\mpl^{-1}$ in which we improved by a factor of $1.5$ between all three resolutions. The convergence of $r\Psi_4$ for low ($\Delta x_{\mathrm{min}}=1.33~\mpl^{-1}$), medium ($\Delta x_{\mathrm{min}}=1.00~\mpl^{-1}$) and high ($\Delta x_{\mathrm{min}}=0.66~\mpl^{-1}$)  resolutions is shown in fig. \ref{fig:Connvergence}.

\section{Abelian Higgs Code Test}\label{sect:Static String}

To test the code, we compare the evolution of a simulation with a known semi-analytic case of the infinite static string \cite{Vilenkin:2000jqa}.
 Given the symmetry of the problem we use polar coordinates
\begin{equation}
\begin{split}
x&=r\cos (\theta )~, \\
y&=r\sin (\theta )~, \\
z&=z~.
\end{split}
\end{equation}
and choose cylindrically symmetric ansatz for the scalar and gauge fields $\phi$ and $A_\mu$
\begin{equation}
\begin{split}
\phi &= f(r)e^{in\theta}~, \\
A_{\theta} &= \frac{n\alpha(r)}{e}~,
\end{split}
\end{equation}
and all other components are set to zero. We impose the boundary conditions
\begin{equation}
\begin{split}
&f(0)=0~, \qquad f(\infty) = 1~, \\ 
&\alpha(0)=0~, \qquad \alpha(\infty) = 1~.
\end{split}
\end{equation}
\begin{figure}[t!]
\begin{center}
{\includegraphics[width=1.0\columnwidth]{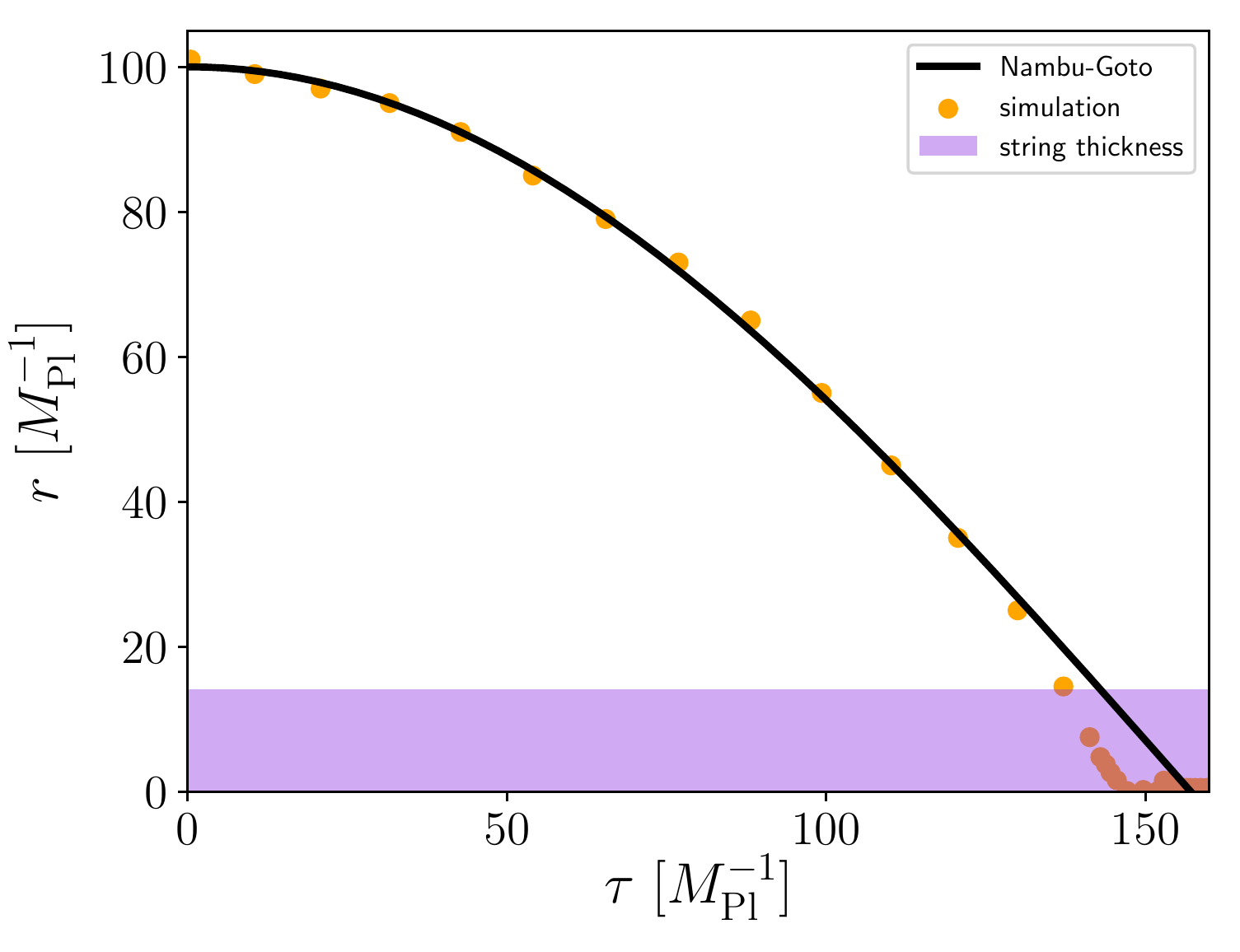}}
\vspace{-1.5em} \caption{\textbf{Comparison with Nambu-Goto} for loop with $G\mu$ = $1.6\times 10^{-2}$ and initial radius $R= 100~\mpl^{-1}$ shows agreement. }
\label{fig:StringPosition}
\end{center}
 \end{figure}
For the metric, the following ansatz is chosen
\begin{equation}
ds^2 = -e^{A(r)}dt^2 + e^{B(r)}(dr^2+r^2d\theta^2) + e^{A(r)}dz^2~,
\end{equation}
where $A(r)$ and $B(r)$ are radial functions numerically determined.  We impose the metric and its derivatives to be locally flat
\begin{equation}\label{eqn:Flat_metric}
\begin{split}
&A(0)=0~, \qquad A^{\prime}(0) = 0~, \\ 
&B(0)=0~, \qquad B^{\prime}(0) = 0~.
\end{split}
\end{equation}

We solve Einstein's and the corresponding matter evolution equations 
\begin{align}
G_{\mu\nu} = 8\pi T_{\mu\nu}~,\label{eqn:EinSteinKG}\\
D_\mu D^\mu \phi = \frac{dV}{d\bar{\phi}} \label{eqn:KG}~,
\end{align}
iteratively as follows.  We solve the Klein-Gordon equation (eq. \ref{eqn:KG}) for fixed flat background, then use this solution to calculate the stress-energy tensor and retrieve the values of $A(r)$ and $B(r)$ via \eqref{eqn:EinSteinKG} to build a new metric. Plugging this back into the Klein-Gordon equation we find new profiles for the fields using the new metric as background. The solution converges quickly (within $\sim 5$ iterations), see fig. \ref{fig:FAProfile} for the obtained profiles of $f$ and $\alpha$.

\section{Comparison with Nambu-Goto} \label{sec:Nambu-Goto}

Previous work showed that without gravity \cite{Nagasawa:1994md} the Nambu-Goto (NG) action is still valid at relativistic speeds.  However, a comparison between the two approaches, leads to consistent results with NG up to roughly the point when the string radius is close to the string thickness  (see fig. \ref{fig:StringPosition}). To reduce gauge effects we use the time of static observer at the position of the string,
\begin{equation}
\tau = \int \alpha|_{\rho = \mathrm{max}(\rho)}~dt~.
\end{equation}
Having shown that NG is a good approximation, we use it to estimate the velocity before unwinding, which we define as the point where the radius of the ring $R$ is equal to the thickness of the string $\delta$. We find 
\begin{equation}
v_\delta = \sqrt{1-\left(\frac{\delta}{R}\right)^2}~,
\end{equation}
which, for our simulations, gives results ranging from $0.97~c$ to $0.99~c$. In the case for which we extract the gravitational wave signal ($G\mu = 1.6\times 10^{-2}$, $R = 100~\mpl^{-1}$) we estimate a velocity of $0.99~c$ before collision.

\bibliography{mybib}
\clearpage 
 
\end{document}